\documentclass[useAMS,usenatbib]{mn2e}

\usepackage{amssymb}
\usepackage{amsmath}
\usepackage{times}
\usepackage{graphicx}

\newcommand{\kms}{\mbox{km s$^{-1}$}}

\newcommand{\apjs}{ApJS}
\newcommand{\apj}{ApJ}
\newcommand{\aj}{AJ}
\newcommand{\araa}{ARA\&A}
\newcommand{\apjl}{ApJL}
\newcommand{\nat}{Nature}
\newcommand{\pasp}{PASP}
\newcommand{\aap}{A\&A}
\newcommand{\mnras}{MNRAS}
\newcommand{\procspie}{Proc. SPIE}

\def\ltsima{$\; \buildrel < \over \sim \;$}
\def\simlt{\lower.5ex\hbox{\ltsima}}
\def\gtsima{$\; \buildrel > \over \sim \;$}
\def\simgt{\lower.5ex\hbox{\gtsima}}

\title[The Gravitational Lens System CASSOWARY\,20]
{CASSOWARY\,20: a Wide Separation Einstein Cross
Identified with the X-shooter Spectrograph\thanks{Based 
on public data from the X-shooter commissioning observations
collected at the European Southern
Observatory VLT/Melipal telescope, Paranal, Chile.}
}

\author[Pettini et al.]{Max Pettini$^1$, 
Lise Christensen$^2$,  Sandro D'Odorico$^{2}$,  
Vasily Belokurov$^1$,  N. Wyn Evans$^1$, 
\newauthor Paul C. Hewett$^1$, Sergey Koposov$^{1,3,4}$, Elena Mason$^5$, and Jo\"{e}l Vernet$^2$\\ 
$^1$ Institute of Astronomy, Madingley Rd, Cambridge, CB3 0HA, UK \\
$^2$ European Southern Observatory, Karl-Schwarzschild-Strasse 2, 85748 Garching bei M\"{u}nchen, Germany\\
$^3$ Max Planck Institute for Astronomy, K\"{o}nigstuhl 17, 69117 Heidelberg, Germany\\
$^4$ Sternberg Astronomical Institute, Universitetskiy prospect, 13, 119992 Moscow, Russia\\
$^5$ European Southern Observatory, Casilla 19001, Santiago 19, Chile\\
}

\begin{document}

\date{Accepted ... Received ... in original form ...}

\pagerange{\pageref{firstpage}--\pageref{lastpage}} \pubyear{2009}

\maketitle

\label{firstpage}

\begin{abstract}
We have used spectra obtained with X-shooter, the triple arm optical-infrared
spectrograph recently commissioned on the
Very Large Telescope (VLT) of the European Southern
Observatory (ESO), to confirm the gravitational lens
nature of the CASSOWARY candidate CSWA\,20. 
This system consists
of a luminous red galaxy at redshift $z_{\rm abs} = 0.741$,
with a very high velocity dispersion, 
$\sigma_{\rm lens} \simeq 500$\,\kms,
which lenses a blue star-forming galaxy at $z_{\rm em}= 1.433$ 
into four images with mean separation of $\sim 6^{\prime\prime}$.
The source shares many of its properties with those
of UV-selected galaxies at $z = 2$--3:
it is forming stars at a rate $\textrm{SFR} \simeq 25 M_\odot$~yr$^{-1}$,
has a metallicity of $\sim 1/4$ solar, and shows nebular emission
from two components 
separated by $0.4^{\prime\prime}$ (in the image plane), 
possibly indicating a merger.
It appears that foreground interstellar material within
the galaxy has been evacuated from the sight-line along which
we observe the starburst, giving 
an unextinguished view of its stars and H\,\textsc{ii} regions.
CSWA\,20, with its massive lensing galaxy producing a high
magnification of an intrinsically luminous background galaxy,
is a promising target for future studies at a variety
of wavelengths.
\end{abstract}


\section{Introduction}
\label{sec:introduction}

The large cosmic volume surveyed by the Sloan Digital Sky Survey (SDSS)
has sparked, among many other projects, several systematic searches
for strong gravitational lens systems (Bolton et al. 2006; Willis et al. 2006;
Estrada et al. 2007; 
Ofek et al. 2008; Shin et al. 2008;
Belokurov et al. 2009; Kubo et al. 2009;
Lin et al. 2009; Wen et al. 2009).
All of these studies share the dual motivation of, on 
the one hand, 
probing the high-mass end of the galaxy mass function
and the underlying distribution of dark matter in the lensing 
galaxies and, on the other, identifying highly magnified high redshift
sources. The latter can bring within reach of current
astronomical instrumentation detailed studies of stellar populations
and interstellar gas at high redshifts which would otherwise have to wait 
until the advent of the next generation of 30+\,m optical-infrared telescopes
(e.g. Pettini et al. 2000, 2002; Teplitz et al. 2000; 
Lemoine-Busserolle et al. 2003; Smail et al. 2007;
Cabanac, Valls-Gabaud, \& Lidman 2008;
Siana et al. 2008, 2009;
Finkelstein et al. 2009; Hainline et al. 2009;
Quider et al. 2009, 2010; Yuan \& Kewley 2009
and references therein).


 \begin{table*}
 \begin{minipage}{132mm}
 \caption{\textsc{SDSS Positions and Magnitudes of Components of CSWA\,20 Lens System}}
   \begin{tabular}{lllccccc} 
\hline 
\hline
Image  & RA (J2000)    & Dec (J2000) & $u$             & $g$             & $r$             & $i$             & $z$              \\ 
\hline 
Lens   &  14~41~49.16  & +14~41~20.6 & $24.6 \pm 1.1$  & $25.1 \pm 0.5$  & $22.7 \pm 0.2$  & $20.70\pm 0.06$ & $20.2 \pm 0.2$   \\
\\
i1     &  14~41~49.38  & +14~41~21.3 & $21.46 \pm 0.15$& $21.60 \pm 0.05$& $21.54 \pm 0.07$& $21.87 \pm 0.17$& $21.76 \pm 0.63$ \\
i2     &  14~41~49.24  & +14~41~22.8 & $21.89 \pm 0.26$& $21.65 \pm 0.06$& $21.40 \pm 0.08$& $21.27 \pm 0.13$& $21.89 \pm 0.88$ \\
i3     &  14~41~48.93  & +14~41~22.2 & $21.56 \pm 0.23$& $21.52 \pm 0.07$& $21.46 \pm 0.11$& $21.27 \pm 0.16$& $21.18 \pm 0.67$ \\
i4     &  14~41~49.18  & +14~41~18.7 & $21.58 \pm 0.23$& $21.57 \pm 0.07$& $21.16 \pm 0.08$& $21.19 \pm 0.16$& $20.59 \pm 0.38$ \\
\\
\hline
     \label{tab:sdss_photom}
 \end{tabular}

Note:  Positional errors are  $\leq 0.1^{\prime\prime}$ for
images i1--i4, and $\simeq 0.1^{\prime\prime}$ for the lensing galaxy.\\

 \end{minipage}
 \end{table*}

The CAmbridge Sloan Survey Of Wide ARcs in the skY 
(CASSOWARY) 
targets multiple, blue companions around massive ellipticals
in the SDSS photometric catalogue as likely candidates
for wide-separation gravitational lens systems. 
A comprehensive description of the search strategy is given by
Belokurov et al. (2009). 
Of the twenty highest priority CASSOWARY candidates,
eight have so far been confirmed as gravitational lenses
and the redshifts of both 
lens and source 
measured---see http://www.ast.cam.ac.uk/research/cassowary/
for further details.

\begin{figure}
\centerline{\hspace{-0.15cm}
\includegraphics[width=0.905\columnwidth,clip,angle=0]{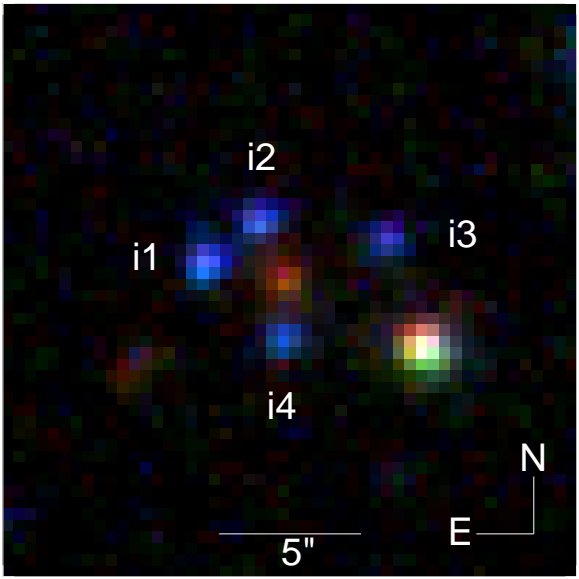}}
\caption{Colour-composite ($g, r , i$)
SDSS image of the CSWA\,20 lens system. Positions and magnitudes of the
different components are listed in Table~\ref{tab:sdss_photom}.}
\label{fig:sdss_image}
\end{figure}

In this paper, we report observations of a ninth 
system, CASSOWARY\,20 or CSWA\,20 for short.
As can be seen from Figure~\ref{fig:sdss_image},
CSWA\,20 consists of four blue images around 
a red galaxy, in a configuration  reminiscent 
of the Einstein Cross (Adam et al. 1989), 
but with a factor of
$\sim 3$ larger separation between the images.
The observations reported here, obtained during the 
commissioning of the triple arm spectrograph X-shooter on the VLT,
confirm the gravitational lens 
nature of the system by showing that three of the four
blue images are of the same source, a star-forming galaxy
at redshift $z_{\rm em} = 1.433$, and that the lens is a massive
elliptical galaxy at $z_{\rm abs} = 0.741$ (the fourth blue image
was not observed, but is most likely to be a fourth gravitationally
lensed image of the $z = 1.433$ galaxy).
Table~\ref{tab:sdss_photom} gives positions and magnitudes 
for the different components of the system. The four images 
of the source are all of similar magnitudes (e.g. $g = 21.52$--21.65)
and all have very blue colours, with $ -0.14 \leq (u - g) \leq +0.24$
and $ +0.06 \leq (g - r) \leq +0.41$. In contrast, the lens has 
red colours with $(g - r) = +2.4$ and $(r - i) = +2.0$.


In Section~2, we give a concise description of the X-shooter instrument 
and details of the observations and data reduction.
Section~3 describes a simple lensing model we have applied to CSWA\,20,
while Sections~4 and 5 present our measurements of the spectra
of the lens and source respectively.
We summarise our results in Section~6.
Throughout this paper we use a 
`737' cosmology with $H_0=70$\,\kms\,Mpc$^{-1}$, 
$\Omega_{\rm M} = 0.3$ and
$\Omega_{\Lambda} = 0.7$.

\section{Observations and Data Reduction}
\label{sec:observations}


 \begin{table*}
 \begin{minipage}{130mm}
 \caption{\textsc{Details of X-shooter Observations of CSWA\,20}}
   \begin{tabular}{llll} 
\hline 
\hline
Date (UT)      & Exp. Time (s)                     & Slit PA ($^\circ$) & Comments            \\ 
\hline  
2009 March 19  & $2 \times 1500$ (UV-B, VIS-R)     & 117                & Two exposures on targets, nodded.\\
               & $2 \times 1600$ (NIR)             &                    & Slit across i1 and i2 (see Figure~\ref{fig:sdss_image}).\\
\\
2009 May 05    & $4 \times 1500$ (UV-B, VIS-R, NIR)& 11                 & Two exposures on targets, two on sky.\\
               &                                   &                    & Slit across i2, lensing galaxy, and marginally i4.\\
\hline
     \label{tab:obs}
 \end{tabular}
 \end{minipage}
 \end{table*}

\subsection{X-shooter}
X-shooter is the first of the second generation VLT instruments 
to be made available to the ESO community; it was built
by a consortium of institutes in Denmark, France, Italy 
and the Netherlands, in collaboration with ESO which
was responsible for the final integration and installation 
on the VLT.
A full description of the instrument is provided 
by D'Odorico et al. (2006).
X-shooter consists of three echelle spectrographs with 
prism cross-dispersion,
mounted on a common structure at the 
Cassegrain focus of the Unit Telescope 2 (Kueyen).
The light beam from the telescope is split in 
the instrument by two dichroics which 
direct the light in the spectral ranges 300--550\,nm and 550-1015\,nm 
to the slits of the UV-B and VIS-R spectrographs respectively. 
The undeviated beam feeds the NIR spectrograph 
with wavelengths in the range 1025--2400\,nm. 
The UV-B and VIS-R spectrographs operate at ambient temperature 
and pressure and deliver two-dimensional spectra on the 
$2048 \times 4102$ 15\,$\mu$m pixel E2V CCD 
and $2048 \times 4096$, 15\,$\mu$m pixel MIT/LL CCD 
respectively.
The NIR spectrograph 
is enclosed in a vacuum vessel and kept at a temperature 
of approximately 80\,K by a 
continuous flow of liquid nitrogen.
The NIR detector is a Teledyne substrate-removed 
HgCdTe Hawaii-2RG array,
with $2048 \times 2048$ 18\,$\mu$m pixels.
The spectral format in the three spectrographs is 
fixed; the final spectral resolution is determined 
by the choice of slit width with each spectrograph 
having its own slit selection device.

\subsection{Observations}
\label{sec:obs}
After two commissioning runs with the UV-B and VIS-R 
spectrographs in November 2008 and January 2009, 
the instrument was operated in its full three-arm configuration
in two further commissioning runs in March 2009 and May 2009 
during which the observations of CSWA\,20 were obtained.
This lens candidate was selected as a good test of the instrument
performance for faint galaxy studies. Different observation strategies
were attempted, as detailed in Table~\ref{tab:obs}. 
In March 2009 the $11^{\prime\prime}$ long
entrance slit was rotated to a position angle on the
sky PA\,$= 117^{\circ}$ and positioned so as to record simultaneously
the spectra of images i1 and i2 (see Figure~\ref{fig:sdss_image}).
In May 2009 the slit was placed across image i2 and the lensing galaxy
(PA\,$= 11^{\circ}$); this set-up also captured some of the light 
from image i4.
For all observations the slit widths were 1$^{\prime\prime}$ (UV-B), 
0.9$^{\prime\prime}$ (VIS-R),
and 0.9$^{\prime\prime}$ (NIR); the corresponding resolving powers
are $R \equiv \lambda/\Delta \lambda = 5100$, 8800, and 5600, 
sampled with 3.2, 3.0, and 4.0 wavelength bins respectively
(after on-chip binning by a factor of two of the UV-B and VIS-R
detectors).  The observations used a nodding along the slit
approach, with an offset of $4^{\prime\prime}$ between 
individual exposures. All calibration frames needed for data processing 
were obtained on the day after the observations.

\subsection{Data Reduction}
\label{sec:redux}

The spectra were processed with a preliminary version of the 
X-shooter data reduction pipeline (Goldoni et al. 2006).
Pixels in the two-dimensional (2D) frames are first mapped to wavelength space
using calibration frames. Sky emission lines are subtracted before 
any resampling using the method developed by Kelson (2003). The different 
orders are then extracted, rectified, wavelength calibrated and merged, 
with a weighted average used in the overlapping regions.  
The final product is a one dimensional, background-subtracted spectrum 
and the corresponding error file.  
Intermediate products including the sky spectrum 
and individual echelle orders are also available.
We followed all of these steps, 
although we used standard \textsc{iraf} tools for 
extracting the 1D spectra (using a predefined aperture)
while this aspect of the pipeline data processing was being
refined.

The VIS-R and NIR spectra were corrected for telluric
absorption by dividing their spectra by that of the O8.5 star 
Hip\,69892, observed with the same instrumental set-up
and at approximately the same airmass as CSWA\,20.
Absolute flux calibration used as reference the
\textit{Hubble Space Telescope} white dwarf standard GD\,71 
(Bohlin et al. 2001) whose spectrum was recorded during the same nights
as CSWA\,20.

\section{Lensing Models}
\label{sec:lensing_model}

As discussed in detail in Sections~\ref{sec:lens} and \ref{sec:source},
the X-shooter spectra show the lens to be an absorption line
galaxy at $z_{\rm abs} = 0.741$, and three of the four
blue images (i1, i2, and i4) to be those of an emission 
line galaxy at  $z_{\rm em} = 1.433$. 
With the assumption that i3 is also a gravitationally lensed 
image of the same galaxy, we can use these redhifts together with the
data in Table~\ref{tab:sdss_photom} to develop some simple lensing
models for CSWA\,20. The aims are to obtain estimates of the enclosed
mass within the images and hence the velocity dispersion of the
lensing galaxy, as well as estimates of the total magnification
of the images.

Before formal modelling, 
let us begin with a very simple model
to estimate rough, order of magnitude effects. 
Suppose the lens is an isolated
singular isothermal sphere with a constant velocity dispersion
$\sigma$. The lensing properties of this model are discussed by
Schneider, Ehlers \& Falco (1992), who show that the typical 
deflection is:
\begin{equation}
\Delta \theta = 1\farcs15 \left( {\sigma_{v} 
\over 200 {\rm \,km\,s}^{-1}}\right)^2 
\left( {D_{\rm ds} \over D_{\rm s}} \right)
\label{eq:isodef}
\end{equation}
where $D_{\rm ds}$ is the angular diameter distance between deflector and source,
whilst $D_{\rm s}$ is the distance between observer and source.  
A simple estimate of the velocity dispersion can be immediately obtained 
by requiring that the isothermal sphere deflection given
by eq.(\ref{eq:isodef})
reproduce the observed deflections ($\Delta \theta \sim 3^{\prime\prime}$)
for the distances $D_{\rm ds}$ and $D_{\rm s}$ implied by 
the values of $z_{\rm abs}$ and $z_{\rm em}$ in our cosmology.
This already suggests that the lensing galaxy is
very massive with $\sigma \sim 500$\,km\,s$^{-1}$.

Of course, mass distributions in nature are not spherically
symmetric. Ellipticity occurs both because of the intrinsic flattening
of the lens galaxy and because of the external tidal shear generated
by neighbouring galaxies. We therefore wish to supplement our simple
model with more realistic, flattened mass distributions. These can
reproduce not just the separations, but also the detailed
positions of all four putative images.


\begin{figure}
\begin{center}
{\hspace{-0.25cm}\includegraphics[height=4.15cm]{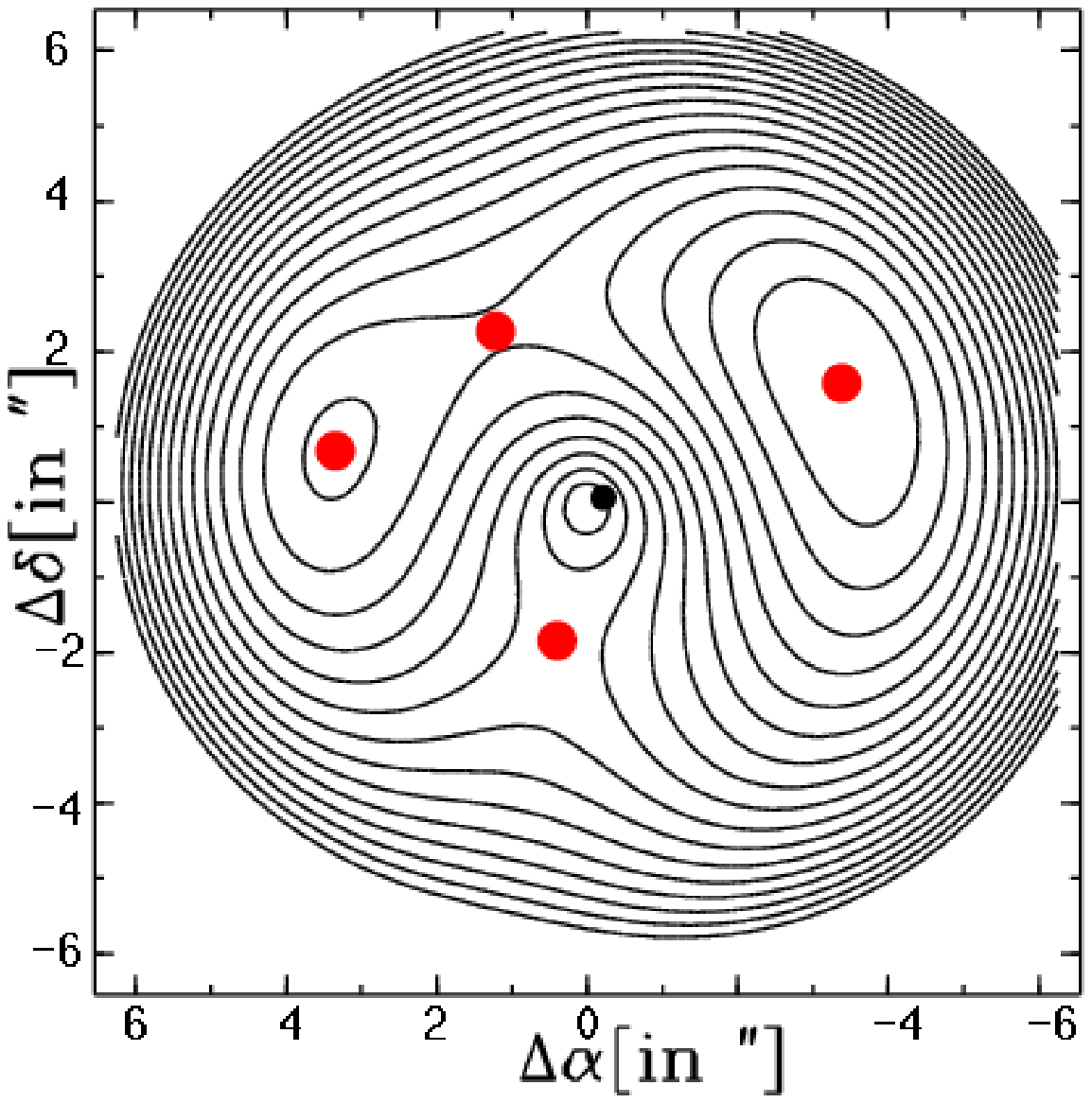}
\includegraphics[height=4.15cm]{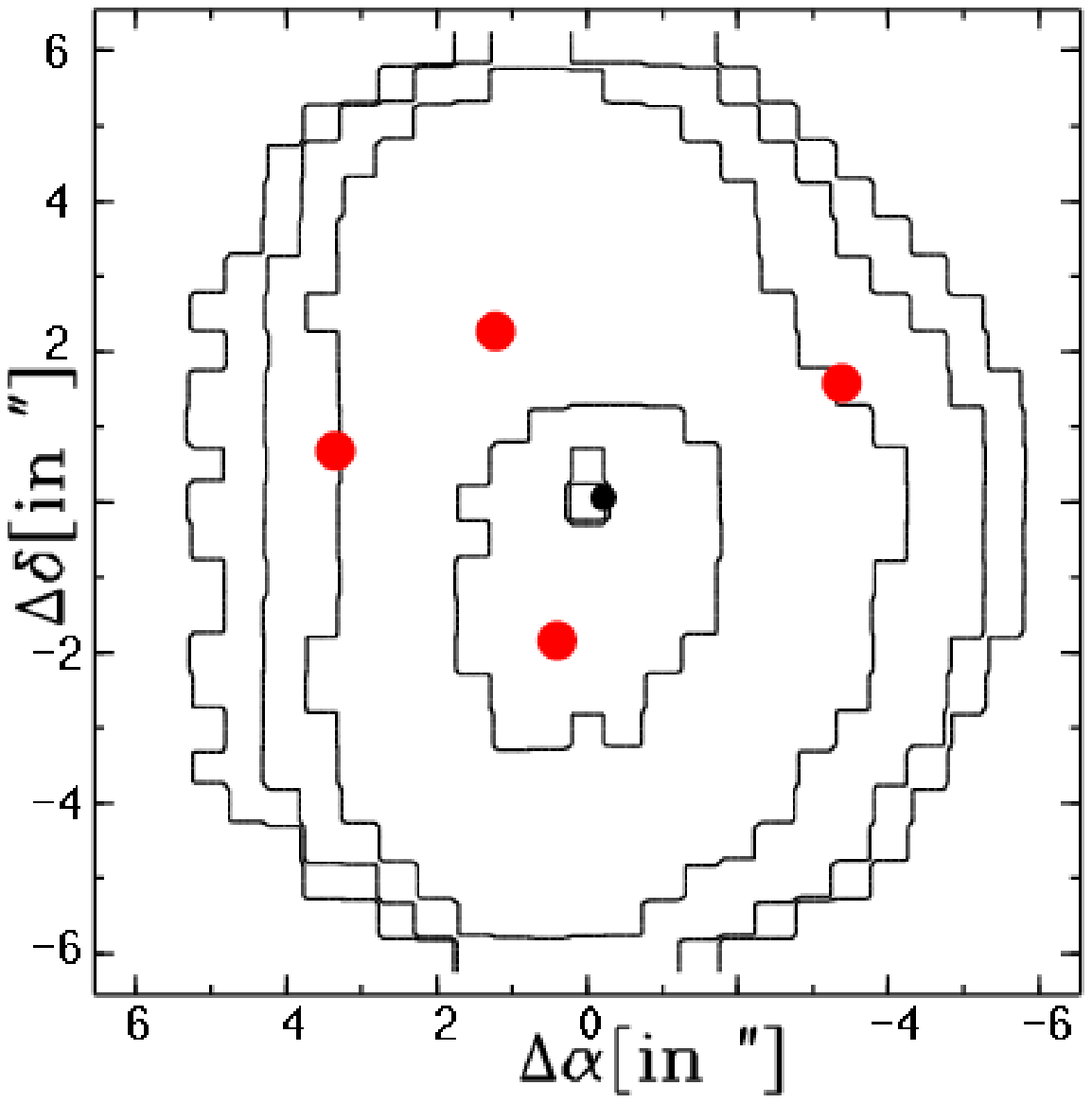}}
\caption{\label{fig:cassmodel} A {\tt PixeLens} model of CSWA\,20 with
  image locations (red dots) and source locations (black dots). The
  left panels shows the Fermat surface, from which the parity of the
  images can be inferred. The right panel shows the pixellated
  mass distribution in units of the critical surface mass density.}
\end{center}
\end{figure}

A powerful way of modelling gravitational lenses is to pixellate the
projected mass distribution of the lensing galaxies into tiles. Mass
can be apportioned to each tile. The mass on the tiles is unknown, but
fixed by requiring that the mass distribution reproduce the images with the
observed parities and locations. Of course the problem is then
under-determined, as there are many more unknowns than
constraints, but can be regularised by requiring that the
mass distribution is isothermal-like.  This simple idea has been
developed by Williams \& Saha (2004) into the {\tt PixeLens} code,
which provides flattened generalisations of the isothermal sphere.

In detail, the solution space for the masses on the tiles is sampled
using a Markov chain Monte Carlo method. We typically generate an
ensemble of 1000 models that reproduce the input data, which in our
case are the image locations 
given in Table~\ref{tab:sdss_photom}.
As the
constraint equations are linear, averaging the ensemble also produces
a solution, which is displayed in Figure~\ref{fig:cassmodel}.


\begin{figure*}
\centerline{\hspace{-0.5cm}
\includegraphics[width=1.9\columnwidth,clip,angle=0]{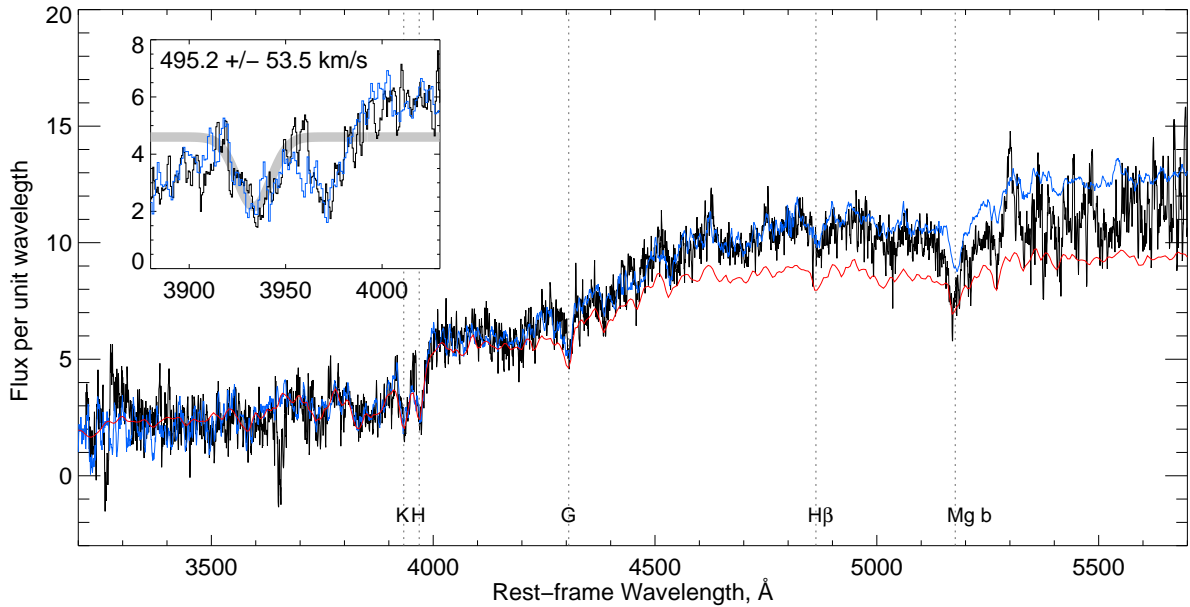}}
\caption{Three spectra are shown in this Figure. 
\textit{Black}: Portion of the 
X-shooter spectrum of the lensing galaxy in CSWA\,20, 
boxcar smoothed with a three-pixel wide filter.
\textit{Blue}: The spectrum of SDSS\,J010354.1+144814.1
the  galaxy with the largest velocity dispersion
($\sigma = 530 \pm 60$\,km~s$^{-1}$)
in the sample of Bernardi et al. (2006).
\textit{Red}: Model spectrum of a 12\,Gyr old single burst
of star formation, from Maraston et al. (2009).
The inset shows a detailed view of the spectral region encompassing
the Ca\,{\sc ii} H \& K lines, together with our Gaussian fit.
}
\label{fig:lens}
\end{figure*}

The left panel shows the arrival time surface. The images lie at the
stationary points of the surface and are marked with red dots. In the
model, we see that the images i1 and i3 correspond to local minima
and so are positive parity, whereas i2 and i4 correspond to
saddles and so are negative parity.  The model predicts an additional,
de-magnified central image which -- as is customary -- is too faint
for detection.  The right panel shows the contours of projected mass
of the absorption line galaxy (in terms of the critical surface mass
density). This is mildly elliptical in the inner parts, and falls off in an
isothermal-like manner.

From the {\tt PixeLens} model, we compute the (cylindrical) mass within
the Einstein radius (21.1\,kpc) as 
$M = (4.4 \pm 0.4) \times 10^{12} M_{\odot}$.
This implies a velocity dispersion for the lensing galaxy of 
$530 \pm 20$\,km~s$^{-1}$, if it is isothermal. 
As we shall see in Section~\ref{sec:lens}, such a high
velocity dispersion is consistent with the widths of the absorption lines
in the X-shooter spectrum of the lens. 
The errors have been calculated
using {\tt PixeLens} to generate 1000 models that reproduce the input
data. From these distributions, we calculate the error on the enclosed
mass by constructing 68 per cent confidence limits. The error on the
enclosed mass gives an error on the velocity dispersion via the Jeans
equations. The total magnification is rather poorly constrained with the
present data, but most of the models have a total magnification (for the
four images) of factors between $\sim 3$ and $\sim 6$

\section{The Lens: a Massive Galaxy at $z_{\rm abs}=0.741$ }
\label{sec:lens}

The spectrum of the deflector galaxy, reproduced in Figure~\ref{fig:lens},
shows the signatures of an old stellar population 
at a redshift $z_{\rm abs} = 0.741$.  
The galaxy has a prominent 4000\,\AA\ break,
and strong Ca\,{\sc ii}~H \& K, G-band and 
Mg$b$\,$\lambda\lambda 5167.3, 5172.7, 5183.6$
absorption features.
H$\beta$ absorption is also present.  
The signal-to-noise ratio of the data
is only modest, but Gaussian fits to the 
Ca\,{\sc ii}~H\&K absorption 
(see inset in Figure~\ref{fig:lens})
result in an
estimate of the velocity dispersion 
$\sigma = 495 \pm 54$\,km~s$^{-1}$ 
after correcting for instrumental broadening
(which is minimal, since $\sigma_{\rm instr} \simeq 15$\,km~s$^{-1}$).  
Additional observations with higher signal-to-noise ratio are
required to provide a significantly improved measure of the velocity 
dispersion.
The overall spectral energy distribution at optical wavelengths
is similar to those of luminous red galaxies (LRGs) at lower redshifts,
as can be appreciated from the comparison in Figure~\ref{fig:lens}
with the SDSS spectrum
of J010354.1+144814.1, the most massive galaxy found by Bernardi et al. (2006)
with no obvious indication that its large velocity dispersion 
($\sigma = 530 \pm 60$\,km~s$^{-1}$) may be due to the superposition
of two objects along the line of sight.
Also shown in Figure~\ref{fig:lens} is the model spectrum computed
by Maraston et al. (2009) for a 12\,Gyr old
single burst of star formation with solar metallicity; this 
comparison further illustrates the very red continuum (longward of the
G-band) of the deflector galaxy in CSWA\,20.

Given the relatively high redshift, and associated $(1+z)^4$ 
surface brightness dimming,
it is not surprising that the galaxy is barely detected in the SDSS
$r$-band image and that the $i$-band magnitude is poorly
determined.
Only the core of the galaxy is evident in the 
SDSS images, with an apparent 
effective radius of $<3$\,kpc, and
deeper, higher-resolution imaging is needed to
determine its structural properties.  However, we can put some
constraints on its total luminosity from the X-shooter spectral data.  The
integrals of the deflector's spectral energy distribution multiplied by the SDSS
$r$-band and $i$-band response curves give $r = 21.7$ and $i = 20.1$ respectively.
Given the slit width of $0.9^{\prime \prime}$, 
a mean seeing of FWHM\,$=0.65^{\prime \prime}$,
and assuming an effective radius of the galaxy $r_{\rm e} = 10$\,kpc
(typical of LRGs, Bernardi et al. 2008)
which corresponds to $r_{\rm e} = 1.1^{\prime \prime}$ 
at $z=0.741$, we calculate that $\sim 40$\% of the total galaxy light falls within the
X-shooter slit.  The estimate of the $i$-band magnitude is then $i =19.1$.
Adopting a $k$-correction of $-1.0$\,mag, an evolutionary correction of 0.8\,mag and
rest-frame colour $(r-i) = 0.4$ (appropriate to a 12\,Gyr old passively evolving
Maraston et al.  2009 model), then gives a corrected $r =18.9$, and a
corresponding absolute
magnitude of $M_{\rm r} = -24.4$. 
With these parameters, this galaxy is among the most massive and luminous 
red galaxies known (Bernardi et al. 2006).

\section{The Source: a Luminous Star-Forming Galaxy at $z_{\rm em} = 1.433$}
\label{sec:source}


\begin{figure*}
\vspace{-0.75cm}
\centerline{\hspace{-0.5cm}
\includegraphics[width=1.4\columnwidth,clip,angle=270]{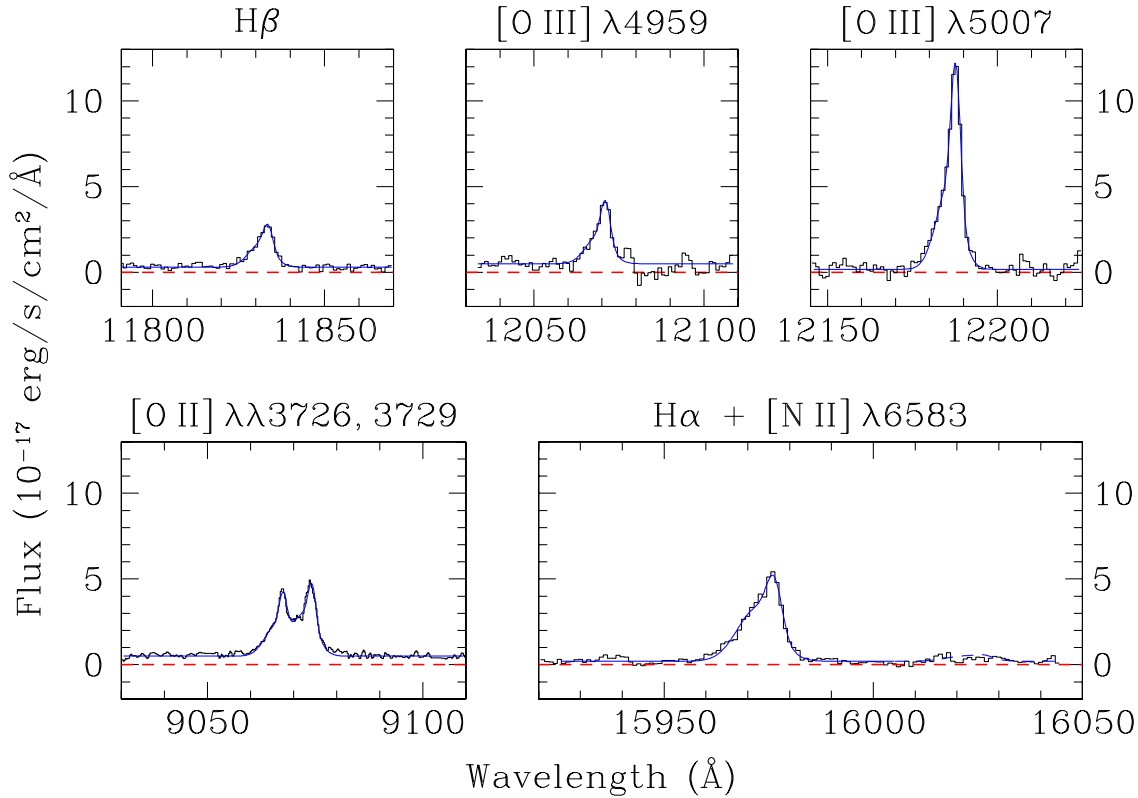}}
\caption{A selection of nebular emission lines in the CSWA\,20 lensed galaxy. 
In each panel, the black histogram shows the data, 
while the blue continuous line is the fit with the parameters 
given in Table~\ref{tab:fits}.
}
\label{fig:em_lines}
\end{figure*}

As mentioned in Section~\ref{sec:obs}, our observations of the source in CSWA\,20
cover images i1 and i2 (the latter observed at two epochs---see Table~\ref{tab:obs}).
In order to improve the signal-to-noise ratio (S/N), 
we added together the spectra of i1 and i2 (after converting the wavelengths
to a vacuum heliocentric frame of reference and binning to a common
wavelength grid) and used this composite spectrum 
in the analysis described below. 
The spectrum shows a number of narrow emission
lines superposed on a weak blue continuum
(see Figure~\ref{fig:em_lines});
the line identifications in Table~\ref{tab:em_lines}
indicate a redshift $z_{\rm em} = 1.433$.

We also recorded image i4 on the second observing run on 2009 May 05,
but the slit position, chosen to cover image i2 and the Lens, only
captured a small fraction of the light of i4. While this spectrum
does show the strongest emission lines at a similar redshift as i1 and i2
(with small differences attributable to the fact that the image
of i4 was offset relative to the slit centre, leading to an offset in
the wavelength calibration), 
it was not included in the composite because
its S/N is much lower than that of i1 and i2.

\subsection{Nebular Emission Lines}
\label{sec:em_lines}

From the lensing model described in Section~\ref{sec:lensing_model},
it was concluded that the overall magnification factor 
for the four images of the Einstein cross is between 
~$\sim 3$ and $\sim 6$. 
It can also be seen from Table~\ref{tab:sdss_photom}
that in the  $u$ and $g$ filters images i1 and i2
account for approximately half of the total flux, that is
$u$(i1+i2)\,$\simeq u$(i3+i4), and similarly for the $g$ magnitudes.
Thus, in all the following analysis we shall assume, for simplicity,
that a magnification factor of $\sim 5/2 = 2.5$ applies to the quantities measured
for i1+i2.

With this assumption and the knowledge of the redshift, 
we can immediately estimate the luminosity of the source,
and compare it with that of other galaxies at similar redshifts.
Luminosity functions at $z \simgt 2$ have been measured mostly
in the rest-frame far-UV continuum, at 1700\,\AA\ 
(e.g. Reddy et al. 2008).
At a redshift $z = 1.433$, this corresponds to an observed
wavelength of 4136\,\AA, which falls between the
transmission of the SDSS $u$ and $g$ filters.
From Table~\ref{tab:sdss_photom} we find that
$u{\rm(i1+i2)} = 20.9$ and $g{\rm(i1+i2)} = 20.87$.
Adopting a magnitude 
$m_{\rm 1700}{\rm (i1+i2)} = 20.9 + 2.5 \log (1+z) = 21.9$ (on the AB scale),
we deduce that at $z = 1.433$ this corresponds to  
an absolute magnitude $M_{\rm 1700}{\rm (i1+i2)} = -23.2$.
Correcting for the magnification by a factor of 2.5,
we find that the source luminosity at 1700\,\AA\ is
$L_{\rm 1700} \simeq 6 L^{\ast}$, compared
to $L_{\rm 1700}^{\ast} = -20.3$ obtained by interpolating
between the luminosity functions of star-forming galaxies
at $z = 1$ and 2 (Reddy \& Steidel 2009).

Such a high luminosity may indicate that our 
lensing model underestimates the magnification of the source.
However, we also note that in the typical $z \sim 2$ galaxy
(equivalent data are not yet available for galaxies at
$z \simeq 1.5$) the UV continuum at 1700\,\AA\ is dimmed
by a factor of $\sim 4-5$ by dust absorption 
(Reddy \& Steidel 2004; Reddy et al. 2006; Erb et al. 2006c).
As we shall
see below, the source in CSWA\,20
is unusual in showing no evidence for dust absorption.
Thus, its high UV luminosity may well be the result
of unusually low reddening, rather than an exceptionally large
number of OB stars.

At $z = 1.433$ our data cover the rest-frame
wavelength interval from $\sim 1350$\,\AA\ to 
$\sim 9050$\,\AA. 
Table~\ref{tab:em_lines} lists the emission lines identified
in the composite X-shooter spectrum of images i1 and i2 of CSWA\,20;
the most prominent of these are reproduced in 
Figure~\ref{fig:em_lines}.
The presence of a number of other spectral features can be
surmised from the data, although their S/N is too low
(with the relatively short integration time devoted to this
object during instrument commissioning) to warrant
their measurement. Among these, we recognise the 
P~Cygni profile of C\,{\sc iv}\,$\lambda 1549.1$
due to massive stars, which is a common feature of
the integrated spectra of star-forming galaxies 
(e.g. Schwartz et al. 2006; Quider et al. 2010).

As can be realised from inspection of Figure~\ref{fig:em_lines},
the line profiles are asymmetric with an extended blue wing,
suggesting that they consist of more than one component.
Inspection of the 2D images confirms the presence of two
components, separated by $\sim 0.4^{\prime\prime}$ along the
slit and by $\sim 100$\,km~s$^{-1}$ in the spectral direction. 
It is possible that the source consists of two merging clumps
of gas and stars.
In future, it should be possible with better data to
extract these two components separately and compare their
properties. Given the limited S/N ratio of the current
observations,  we opted for a single extraction 
(Section~\ref{sec:redux})
which blends together the light from the two clumps,
and results in the asymmetric line profiles evident in 
Figure~\ref{fig:em_lines}.
When these emission lines are analysed with Gaussian fitting routines,
as explained below, they appear to consist of two
components with the parameters listed in Table~\ref{tab:fits}.
However, it is important to keep in mind
that these parameters refer to the \textit{blend}
of the emission lines from the two clumps, and may well
turn out to be different from the individual values
appropriate to each clump when the two are analysed separately. 
However, the \textit{total} flux values deduced for the emission
lines and listed in Table~\ref{tab:em_lines} should not
be affected.

Thus, in order to deduce redshifts $z_{\rm em}$, velocity dispersions $\sigma$,
and line fluxes $F$, we fitted the emission lines with two
Gaussian components, using ELF (Emission Line Fitting) 
routines in the Starlink \textsc{dipso}
data analysis package (Howarth et al. 2004), 
as well as custom-built software.
The fitting proceeded as follows.


 \begin{table}
 \begin{minipage}{80mm}
 \caption{\textsc{Emission Lines identified in Images i1 and i2 of CSWA\,20}}
   \begin{tabular}{llll} 
\hline 
\hline
Line            & $\lambda_{\rm lab}^{\rm a}$ (\AA)   & $F^{\rm b}$  & Comments            \\ 
\hline  
[N\,{\sc ii}]   & 6585.27                             & $\leq 0.45$          & This line is undetected \\

H$\alpha$       & 6564.614                            & $4.9 \pm 0.1$        & Affected by sky residuals \\

[O\,{\sc iii}]  & 5008.239                            & $7.0 \pm 0.1$        & \\              

[O\,{\sc iii}]  & 4960.295                            & $2.23 \pm 0.07$      & \\

H$\beta$        & 4862.721                            & $1.65 \pm 0.04$      & \\

[O\,{\sc ii}]   & 3729.86                             & $2.09 \pm 0.03$      & Blended with [O\,{\sc ii}]\,$\lambda 3727$ \\

[O\,{\sc ii}]   & 3727.10                             & $1.56 \pm 0.03$      & Blended with [O\,{\sc ii}]\,$\lambda 3729$ \\

Mg\,{\sc ii}    & 2803.5324                           & $0.27 \pm 0.05$      & \\
 
Mg\,{\sc ii}    & 2796.3553                           & $0.48 \pm 0.06$      & \\

C\,{\sc iii}]   & 1908.734                            & ($0.18$)$^{\rm c}$              & Noisy \\

[C\, {\sc iii}]  & 1906.683                           & ($0.25$)$^{\rm c}$              & Noisy \\
\hline

 \end{tabular}
$^{\rm a}$ Vacuum wavelengths.\\
$^{\rm b}$ Integrated line flux in units of 10$^{-16}$\,erg~s$^{-1}$~cm$^{-2}$.\\
$^{\rm c}$ Uncertain measurement.\\
     \label{tab:em_lines}
\end{minipage}
\end{table}

The best observed emission lines among those listed in 
Table~\ref{tab:em_lines} are the [O\,\textsc{iii}] doublet
and H$\beta$, as they
are relatively strong and free from sky residuals.
Consequently, we first fitted these three nebular lines,
using custom-built Gaussian fitting routines which allowed
the redshift, velocity dispersion, and flux in each component
to vary but with the constraint that the redshift 
and velocity dispersion of each
component should be the same for all three lines,
and therefore using the data from all three lines
to find the values of $z_{\rm em}$ and $\sigma$
which minimise the difference between computed and observed
profiles. Errors on the values of $z_{\rm em}$, $\sigma$,
and $F$ so deduced were estimated using a Monte Carlo approach,
whereby the best fitting computed profile was perturbed with a 
random realisation of the error spectrum and refitted. The process
was repeated 100 times and the error in each quantity 
($\delta z_{\rm em}$, $\delta \sigma$, $\delta F$)
taken to be the standard deviation of the values generated by the 
100 Monte Carlo runs.

Table~\ref{tab:fits} lists the values of $z_{\rm em}$, and $\sigma$
so derived (after subtracting in quadrature the value of
$\sigma_{\rm INSTR}$ corresponding to the instrumental
resolution of the appropriate X-shooter spectrum---see
Section~\ref{sec:obs}). 
The theoretical profiles computed with the
parameters in Table~\ref{tab:fits} and the line fluxes in Table~\ref{tab:em_lines}
are shown with continuous lines in Figure~\ref{fig:em_lines}.

In the next stage of the process we applied the values of $z_{\rm em}$ and $\sigma$
deduced from the analysis of the [O\,\textsc{iii}] and H$\beta$ lines to the 
[O\,\textsc{ii}] doublet which is only partially resolved, leaving the flux
in each member of the doublet as the only free parameter. As can be seen 
from Figure~\ref{fig:em_lines}, the model parameters in Table~\ref{tab:fits}
provide a satisfactory fit to the [O\,\textsc{ii}] doublet.
The fit to H$\alpha$ seemingly required a small shift in the 
redshift of the narrower component, corresponding to a velocity
difference $\Delta v = +25$\,km~s$^{-1}$, or $\sim2$ wavelength bins,
but we suspect that this is an artifact of sky residuals which
affect the H$\alpha$ emission line more than other spectral features.
The integrated flux in the line is the same independently of whether
this shift is applied or not. 

The measurements collected in Tables~\ref{tab:em_lines} and \ref{tab:fits}
allow us to determine a range of physical properties for the lensed
galaxy in CSWR\,20. Before discussing the more involved derivations,
we point out two straightforward conclusions. First, we note that
the ratio F(H$\alpha$)/F(H$\beta) = 2.95 \pm 0.1$ is as expected
from Case B recombination, F(H$\alpha$)/F(H$\beta) = 2.86$
(Brocklehurst 1971),
with the `standard' parameters
of temperature $T = 10\,000$\,K and
electron density $n(e) = 100$\,cm$^{-3}$
(the density dependence is in any case minimal, and the temperature
dependence is minor).
Thus it appears that the emission line gas is essentially 
unreddened.
A lack of dust in this galaxy (as viewed from Earth) is further
indicated by the blue UV continuum: the photometry in Table~\ref{tab:sdss_photom}
implies a UV spectrum which is flat in $f_{\rm \nu}$, as 
expected for an unreddened population of OB stars
(e.g. Bruzual \& Charlot 2003).\footnote{Galactic
extinction is negligible in this direction, 
with $E(B-V)_{\rm MW} = 0.022$ (Schlegel, Finkbeiner, \& Davis 1998).} 

Second, the [O\,\textsc {ii}] doublet ratio, which is
sensitive to density (Osterbrock 1989), is found to be close
to the low density limit: $F(3729)/F(3727) = 1.33$ implies
$n(e) = 110$\,cm$^{-3}$. 
While at face value the ratio of the [C\,\textsc{iii}] doublet lines
would suggest much higher electron densities
($> 1000$\,cm$^{-3}$),
these lines are considerably weaker than [O\,\textsc{ii}]
and their ratio is much more uncertain, so that we consider
the value of $n(e)$ deduced from [O\,\textsc{ii}] to be the
more reliable.



 \begin{table}
 \begin{minipage}{100mm}
 \caption{\textsc{Profile Decomposition of [O\,\textsc{iii}] and H$\beta$ Emission Lines}}
   \begin{tabular}{cccc} 
\hline 
\hline
~~~~~~Component~~~~~~       & ~~~~~~$z_{\rm em}$~~~~~~~           & ~~~~~~$\sigma^{\rm a}$ (km~s$^{-1}$)~~~~~~   &\\ 
\hline  
1               & $1.43308 \pm 0.00004$  & $85 \pm 3$               & \\

2               & $1.43354 \pm 0.00001$  & $26 \pm 3$               & \\
\hline
 \end{tabular}

$^{\rm a}$ Corrected for the instrumental resolution.\\

     \label{tab:fits}
\end{minipage}
\end{table}

\subsection{Star-Formation Rate}

The H$\beta$ flux given in Table~\ref{tab:em_lines} implies 
a luminosity $L({\rm H}\beta) = 2.1 \times 10^{42}$\,erg~s$^{-1}$
in our cosmology. 
Adopting the Case B recombination 
ratio\footnote{We go through this route,
rather than using the H$\alpha$ flux directly, because
the H$\alpha$ emission line is contaminated by strong
sky line residuals and we consider the measurement of its flux
less reliable than that of H$\beta$. However, in practice
we would have obtained essentially the same result had we adopted
the value of $F$(H$\alpha$) from Table~\ref{tab:em_lines}
as our starting point in the following calculation.},
F(H$\alpha$)/F(H$\beta) = 2.86$, the corresponding H$\alpha$
luminosity is 
$L({\rm H}\alpha) = 6.0 \times 10^{42}$\,erg~s$^{-1}$,
which in turn implies a star formation rate:
\begin{equation}
{\rm SFR}  = 7.9 \times 10^{-42}  L({\rm H}\alpha) 
\cdot \frac{1}{1.8} \cdot \frac{1}{2.5} \cdot 2
= 21 (M_\odot \, {\rm yr}^{-1}) ~ .
\label{eq:sfr_ha}
\end{equation}

The first term on the right-hand side of eq.~(\ref{eq:sfr_ha})
is the conversion factor between $L$(H$\alpha$) and SFR
proposed by Kennicutt (1998),  to which we apply three corrections, 
as follows. 
The first adjustment, by the factor of 1/1.8, takes into account
the flattening of the stellar initial mass function (IMF) 
for masses below $1 M_\odot$ (Chabrier 2003)
compared to the single power
law of the Salpeter IMF assumed by Kennicutt (1998).
The second correction factor is the $2.5\times$ magnification
we estimated for the sum of images i1 and i2 (Section~\ref{sec:lensing_model}).
The last term corrects for light loss through the spectrograph slit,
which we estimate to be a factor of $\sim 2$ 
by comparing the measured UV flux in the
X-shooter spectrum with the SDSS magnitudes given in Table~\ref{tab:sdss_photom}.
A factor of $\sim 2$ slit loss is typical of near-IR observations
of nebular emission lines from high-$z$ galaxies (e.g. Erb et al. 2006c).

An independent measure of the SFR is provided by the
UV continuum from OB stars. 
From $u{\rm(i1+i2)} = g{\rm(i1+i2)} = 20.9$
(Table~\ref{tab:sdss_photom}), we have 
$f_\nu(1700) =  1.6 \times 10^{-28}$\,erg~s$^{-1}$~cm$^{-2}$~Hz$^{-1}$
from the definition of AB magnitudes in terms of $f_\nu$.
The corresponding luminosity 
$L_\nu(1700) = 8.3 \times 10^{29}$\,erg~s$^{-1}$~Hz$^{-1}$
in turn implies:
\begin{equation}
{\rm SFR}  = 1.4 \times 10^{-28}  L_\nu(1700)
\cdot \frac{1}{1.8} \cdot \frac{1}{2.5} 
= 26 (M_\odot \, {\rm yr}^{-1}) ~ ,
\label{eq:sfr_uv}
\end{equation}
using Kennicutt's (1998) scaling between $L_{\rm UV}$ and SFR
and applying the same corrections as above for the Chabrier (2003)
IMF and magnification factor.
The good agreement between the SFR estimates from the UV continuum
and the Balmer lines is a further indication that the OB stars and H\,{\sc ii}
regions of this galaxy suffer very little reddening from 
dust.

\subsection{Metallicity}

Since we detect emission lines of [O\,\textsc{ii}],  
[O\,\textsc{iii}],  and H$\beta$, we can use the 
$R_{23} \equiv [F(3726) + F(3729) + F(4959) + F(5007)]/F({\rm H}\beta)$ 
index first proposed by Pagel et al. (1979) 
as an approximate measure of the oxygen abundance
in the absence of direct temperature diagnostics.
Over the thirty years since the seminal
paper by Pagel and collaborators, many studies have shown that
the values of (O/H) so deduced are accurate to 
within a factor of $\sim 2$ when applied to the integrated
spectra of galaxies (e.g. Pettini 2006; Kewley \& Ellison 2008
and references therein).

\begin{figure}
\vspace{-2.5cm}
\centerline{\hspace{0.5cm}
\includegraphics[width=1.25\columnwidth,clip,angle=270]{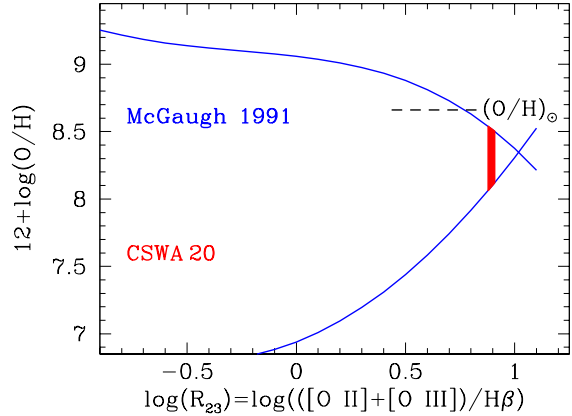}}
\vspace{-1.75cm}
\caption{Oxygen abundance from the $R_{\rm 23}$ index. 
The continuous lines are the calibration byMcGaugh (1991) for the measured
ionization index $O_{\rm 32} = $\,[O\,\textsc{iii}]/[O\,\textsc{ii}],
while the shaded area shows the values allowed by the measured 
$R_{\rm 23}$ line ratio and its statistical $1 \sigma$ error. 
The broken horizontal line is the reference solar oxygen
abundance $12 + \log {\rm (O/H)} = 8.69$ (Asplund et al. 2009).
}
\label{fig:r23}
\end{figure}

Figure~\ref{fig:r23} shows the dependence of (O/H)
on $R_{23}$ using the analytical 
expressions by McGaugh (1991) as given by
Kobulnicky et al. (1999); 
these formulae express O/H in terms of
$R_{23}$ and the ionization index 
O$_{32} \equiv [F(4959) + F(5007)]/[F(3726) + F(3729)]$.
The double-valued nature of the $R_{23}$ index
translates to an uncertainty in the oxygen abundance
between $12 + {\rm (O/H)} = 8.09$ and 8.52,
or between $\sim 1/4$ and $\sim 3/4$ of the solar abundance
$12 + {\rm (O/H)}_\odot = 8.69$ 
(Asplund et al. 2009).

The degeneracy can be broken by considering the 
[N\,\textsc{ii}]/H$\alpha$ ratio. 
According to Pettini \& Pagel (2004),
our upper limit on
the $N2 \equiv \log [F(6583)/F(H\alpha)]$
index, $N2 \leq -1.04$ (Table~\ref{tab:em_lines}), implies 
$12 + {\rm (O/H)} \leq 8.26$, or $\leq 2/5 {\rm (O/H)}_{\odot}$,
favouring the lower branch solution of the $R_{23}$ method.
A metallicity of $\sim 1/4$ solar is not unusual for
star-forming galaxies at $z =  2$--3 (Erb et al. 2006a; Maiolino et al. 2008).


\subsection{Mg\,{\sc ii} Emission}

We conclude this Section by commenting briefly 
on the presence of narrow Mg\,{\textsc{ii}\,$\lambda\lambda 2796, 2803$
lines among the nebular emission seen in CSWA\,20.
These lines are weak (see Figure~\ref{fig:MgII}), although undoubtedly
real ($8 \sigma$ and $5 \sigma$ detections 
respectively---see Table~\ref{tab:em_lines}).
To our knowledge, narrow Mg\,\textsc{ii} emission
has rarely been reported in nearby extragalactic
H\,\textsc{ii} regions, but this could be due to the fact that
this wavelength region has not been observed 
extensively in extragalactic nebulae and, if the two lines
are generally weak, existing observations  
may not have the sensitivity and resolution
to detect them. 

\textit{Broad} Mg\,\textsc{ii} emission
is of course a common feature in the spectra of Active Galactic
Nuclei (AGN), but we find no evidence in our spectrum of CSWA\,20
for the presence of an AGN.
Specifically: (a) the Mg\,\textsc{ii} lines are narrow
(the profile decomposition into two Gaussians returns values
of $\sigma$ comparable to those listed in Table~\ref{tab:fits});
(b) we detect no high ionisation emission lines due to an AGN
in the rest-frame spectral range 1350--9050\,\AA\ of our
data; and (c) the nebular emission line ratios fall well away
from the locus occupied by AGN in diagnostic diagrams
such as that shown in Figure~\ref{fig:bpt}.

It is possible that in star-forming galaxies at intermediate
redshifts weak Mg\,{\sc ii} emission is more common
than anticipated. In their survey of 1406 galaxies at $z \sim 1.4$,
Weiner et al. (2009) identified a small proportion of galaxies
(50/1406, or $\approx 3.5$\%) with `excess' Mg\,{\sc ii} emission
whose nature they were unable to establish with certainty.
However, even their stacked spectrum of the remaining 
1356 galaxies does show weak Mg\,{\sc ii} P Cygni profiles,
with broad blueshifted absorption and weak emission redshifted
by a few 10s of km~s$^{-1}$. Galaxies with `excess' emission
tend to be bluer than average.

CSWA\,20 appears to fit into this general pattern, with the 
exception that no strong absorption component is evident. 
Thus, it is
possible that the unusual lack of absorbing material
in front of the stars and H\,\textsc{ii} regions
suggested by the negligible reddening noted above
gives an unimpeded view of the intrinsic Mg\,{\sc ii}
emission from the H\,{\sc ii} regions of this galaxy.
It is interesting that, when fitted with two Gaussian 
components, as described
in Section~\ref{sec:em_lines} for the other nebular lines,
the best fitting values of redshift for the two components in Mg\,{\sc ii}
differ by $\Delta v = +30$\,km~s$^{-1}$ ($\sim 3$ wavelength bins)
from the corresponding values listed in Table~\ref{tab:fits}.
We do not believe that this difference is due to an incorrect
wavelength calibration, because we checked the accuracy 
of the wavelength scale
in this region of the spectrum by reference to sky emission
lines, and found it to be $\pm 2$\,km~s$^{-1}$ 
($\sim 1/5$ of a wavelength bin).
If the shift to longer wavelengths is not due to noise, it may be an 
indication of radiation transfer effects 
in an expanding medium, analogous to those
commonly seen in the Lyman~$\alpha$ line
(e.g. Verhamme, Schaerer, \& Maselli 2006;
Steidel et al. 2010).  


\begin{figure}
\vspace{-1.5cm}
\centerline{\hspace{1.15cm}
\includegraphics[width=1.25\columnwidth,clip,angle=270]{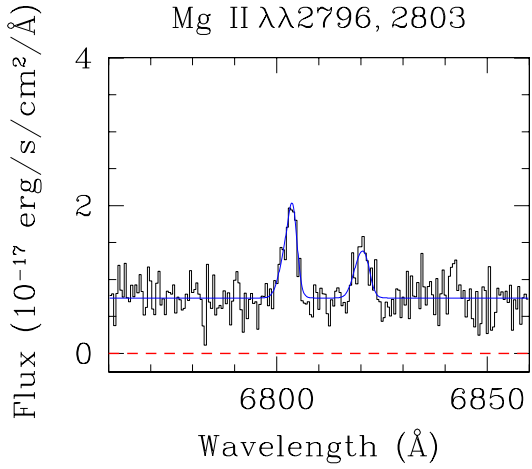}}
\vspace{-1.75cm}
\caption{Mg\,\textsc{ii}\,$\lambda\lambda 2796, 2803$ emission lines 
in the CSWA\,20 lensed galaxy. 
The black histogram shows the data, 
while the blue continuous line is our
two component Gaussian fit (see text for further details).
}
\label{fig:MgII}
\end{figure}

\begin{figure}
\vspace{-3.75cm}
\centerline{\hspace{0.25cm}
\includegraphics[width=1.35\columnwidth,clip,angle=0]{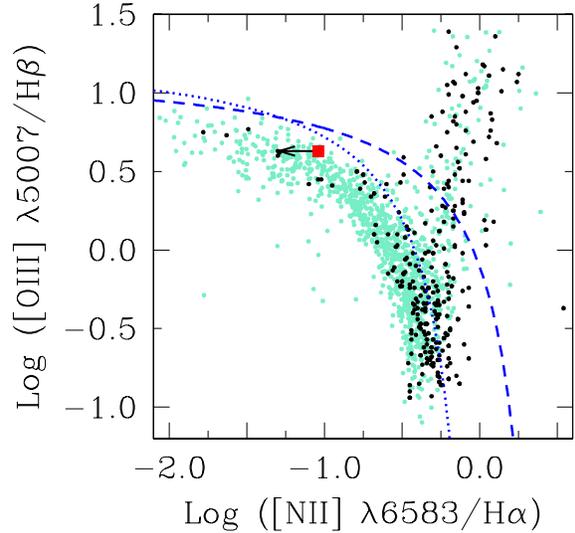}}
\vspace{-3.3cm}
\caption{[O\,\textsc{iii}]/H$\beta$ vs. [N\,\textsc{ii}]/H$\alpha$ diagnostic diagram.
The location of the lensed galaxy in CSWA\,20 is 
shown by the red square and left-pointing arrow. The small green circles
are galaxies from the KISS survey (Salzer et al. 2005) and the small black
dots are local starburst galaxies from Kewley et al. (2001). The dashed line
shows the locus of points which Kewley et al. (2001) consider to be the
theoretical limit for starbursts, in the sense that galaxies without an AGN
component should fall below and to the left of this line. The dotted line is
an empirical determination by Kauffmann et al. (2003) of the same limit. 
The presence of an AGN in CSWA\,20 seems unlikely, given the weakness
of the  [N\,\textsc{ii}] emission. 
}
\label{fig:bpt}
\end{figure}

\section{Summary and Conclusions}

We have presented X-shooter observations, which are
among the first obtained with this new VLT instrument,
confirming the gravitational lens nature of 
CSWA\,20. This system,
originally identified as a candidate from
the CASSOWARY search of SDSS images,
is found to consist of a luminous red galaxy
at $z_{\rm abs} = 0.741$ which magnifies the light
from a background star-forming galaxy at $z _{\rm em} = 1.433$
into four images of approximately equal brightness.
With a velocity dispersion 
$\sigma_{\rm lens} \simeq 500$\,km~s$^{-1}$,
the lensing galaxy is among the most massive known;
the mass $M \sim 4 \times 10^{12} M_\odot$ enclosed
within the Einstein radius ($\sim 21$\,kpc)
is responsible for the wide separations
($\simeq 6^{\prime\prime}$) between the four
gravitationally lensed images of the source.

The source blue colours  satisfy the `BM' photometric
selection criteria of Steidel et al. (2004), which isolate galaxies 
in the redshift interval $1 .4 \simlt z \simlt 2.0$.
BM galaxies have been relatively little studied up to now.
The source in CSWA\,20 shares many of the properties
of the much better characterised sample of `BX' galaxies
at $z \simeq 2$. Its star-formation rate,
$\textrm{SFR} \simeq 25 M_\odot$~yr$^{-1}$,
metallicity $\textrm{(O/H)} \simeq 0.25\, \textrm{(O/H)}_\odot$,
and velocity dispersion (equivalent to two components
separated by $\Delta v = 57$\,\kms, with 
$\sigma_1 = 85$\,\kms\ and $\sigma_2 = 26$\,\kms),
are all within the range of values appropriate to
star-forming galaxies at $z = 2$--3. 
The galaxy in CSWA\,20 is, however, unusual in its low degree
of reddening; only $\sim 5$\% of the BX galaxies
studied by Erb et al. (2006b) have values of 
$E(B-V) \simeq 0$, as found for this source.
Evidently, we view the starburst
from a direction along which most of the foreground
interstellar material in the galaxy has been evacuated.
The lack of dust extinction probably contributes to 
the unusually high UV luminosity of this galaxy, 
$L_{1700\,\AA} \sim 6 L^\ast_{1700\,\AA}$,
and the absence of absorbing gas in front of the H\,\textsc{ii}
regions may be the reason why we can detect weak 
Mg\,\textsc{ii} resonance lines in \textit{emission}.

The observations presented here are a good demonstration
of the new opportunities for studies of high-$z$ galaxies
afforded by the availability of X-shooter. 
Data secured with less
than two hours of integration on two of the four images
(with magnitudes $g \simeq r \simeq 21.5$) have been
sufficient to characterise many of the salient properties
of both lens and source, thanks to the high efficiency
and wide wavelength coverage of this unique instrument.
Longer observations, achieving higher
S/N ratios, will in future shed further light on the
nature of both the massive foreground LRG and the background
star-forming galaxy. The gravitational magnification 
will make it possible to study the internal structure of the latter
on unusually small scales, by comparing  the individual spectra of
the four images (e.g. Stark et al. 2008). 
With more strongly-lensed systems continuing to be found,
the characteristics of X-shooter will veritably open a 
new window on the high-redshift universe, well ahead
of the era of 30+\,m telescopes.

\section*{Acknowledgements}
The good quality of the spectra obtained during 
the commissioning runs of the instrument 
was the result of the dedicated and successful efforts by
the entire X-shooter Consortium team. 
More than 60 engineers, technicians, and astronomes worked
over more than five years on the project in Denmark, France, Italy, 
the Netherlands, and at ESO. 
S.D. would like to acknowledge, in representation of the whole team, 
the co-Principal Investigators 
P.~Kjaergaard-Rasmussen, F.~Hammer, R.~Pallavicini, L.~Kaper, and 
S.~Randich, and the Project Managers
H.~Dekker, I.~Guinouard, R.~Navarro, and F.~Zerbi. 
Special thanks also go to the ESO commissioning team, 
in particular H.~Dekker,  J.~Lizon, R.~Castillo, M.~Downing, 
G.~Finger, G.~Fischer, C.~Lucuix, P.~Di~Marcantonio,
A.~Modigliani, S.~Ramsay and P.~Santin. 
Finally, we are grateful to Dan Nestor for kindly sharing his Gaussian
decomposition code with us, to Lindsay King, 
Jason Prochaska, Naveen Reddy and Alice Shapley
for illuminating discussions, 
and to the anonymous referee whose comments
and suggestions improved the paper.
SK was supported by the DFG through SFB~439, by
a EARA-EST Marie Curie Visiting fellowship and partially
by RFBR 08-02-00381-a grant.

\label{lastpage}

\end{document}